\documentstyle[12pt]{article}
\newcommand{\newsection}{ \setcounter{equation}{0} \section}
\def\appendix#1{
  \addtocounter{section}{1}
  \setcounter{equation}{0}
  \renewcommand{\thesection}{\Alph{section}}
  \section*{Appendix \thesection\protect\indent #1}
  \addcontentsline{toc}{section}{Appendix \thesection\ \ \ #1}
  }
\newcommand{\beq}{\begin{equation}}
\newcommand{\eeq}{\end{equation}}
\newcommand{\bea}{\begin{eqnarray}}
\newcommand{\eea}{\end{eqnarray}}
\newcommand{\}{\`a}
\newcommand{\bbm}{Baxter-Bazhanov model}
\newcommand{\bbmm}{Baxter-Bazhanov Model}
\newcommand{\bg}{braid group}
\newcommand{\zbn}{\mbox{\hbox{\bf Z}$_N$}}
\newcommand{\zsn}{\mbox{\hbox{\bf Z}$_n$}}
\newcommand{\complex}{\hbox{\bf C}}
\newcommand{\tr}{\hbox{\rm Tr}\;}
\newcommand{\zet}{\hbox{\bf Z}}

\newcommand{\cala}{{\cal A}}
\newcommand{\calt}{{\cal T}}
\newcommand{\calm}{{\cal M}}
\newcommand{\call}{{\cal L}}

\newcommand{\calf}{{\cal F}}
\newcommand{\calr}{{\cal R}}
\newcommand{\calw}{{\cal W}}
\newcommand{\calp}{{\cal P}}

\begin{document}
\topmargin 0pt
\oddsidemargin 5mm
\headheight 0pt
\headsep 0pt
\topskip 9mm

\addtolength{\baselineskip}{0.20\baselineskip}

\hfill IFUM - 465/FT

\hfill March 1994
\begin{center}

\vspace{30pt}
{\large \bf
 Three-Dimensional Integrable Models\\
 and Associated Tangle Invariants
}

\vspace{30pt}
{\sl B.~L.~Cerchiai}

\vspace{10pt}
{\footnotesize
Dipartimento di Fisica, Universit\ di Milano,\\
via Celoria 16, 20133 Milano, Italy\\
}

\vspace{20pt}

{\sl M.~Martellini}

\vspace{10pt}
{\footnotesize
Dipartimento di Fisica, Universit\ di Milano,\\
via Celoria 16, 20133 Milano, Italy\\
and INFN, Sezione di Pavia, 27100 Pavia, Italy \\
E.mail: {\it martellini@vaxmi.mi.infn.it}}

\vspace{20pt}

and

\vspace{20pt}

{\sl F.~Valz-Gris}

\vspace{10pt}
{\footnotesize
Dipartimento di Fisica, Universit\ di Milano,\\
via Celoria 16, 20133 Milano, Italy\\
E.mail: {\it valzgris@mite35.mi.infn.it}}

\end{center}

\vspace{20pt}

\vfill

\begin{center}
{\bf Abstract}
\end{center}
In this paper we show
that the Boltzmann weights of the three-dimensional
{\bbm} give representations of the \bg,
if some suitable spectral limits are taken.
In the trigonometric case we classify all possible spectral limits
which produce {\bg} representations.
Furthermore we prove that for some of them we get cyclotomic
invariants of links and for others we obtain tangle invariants
generalizing the cyclotomic ones.

\vfill

\newpage

\topskip 4mm

\newsection{Introduction}

In paper \cite{baxbaz1} Baxter and Bazhanov introduced
a particularly interesting three-di\-mensional integrable model
with N local states. It is one of the few solvable
three-dimensional models and seems to be highly non-trivial.

The {\bbm} is a generalization of the Zamolodchikov model~\cite{zamo1,zamo2},
which is the particular case N=2.
Kashaev, Mangazeev and Stroganov~\cite{kash1,kash2} proved that
the Boltzmann weights of the {\bbm}
satisfy the tetrahedron equations~\cite{zamo1,jaekel,stroganov}.
This is a generalization of the result obtained by Baxter in~\cite{baxter}
for the Zamolodchikov model.
They use the symmetry properties~\cite{kash1} of the Boltzmann weights,
which have been found independently also
by Baxter and Bazhanov~\cite{baxbaz2}.

One of the most important features~\cite{baxbaz1} of the {\bbm}
is that apart from a modification of the boundary conditions it can
be obtained as a three-dimensional interpretation of the
generalized sl(n)-chiral Potts model~\cite{bkms,djmm1,nakanishi}.

Given a two-dimensional integrable model, which has Boltzmann weights
satisfying the Yang-Baxter equation, it is an interesting question to
ask which {\bg} representations and hence which link invariants
arise therefrom. Akutsu,Deguchi and Wadati~\cite{aku1,aku2}
invented a general procedure to study this problem
and obtained link invariants from most two-dimensional
integrable models. Date, Jimbo, Miki and Miwa~\cite{djmm2} studied
the {\bg} representations and the corresponding (cyclotomic) link invariants
arising from the sl(n)-chiral Potts model in the trigonometric limit.
Following a suggestion made by Jones~\cite{jones}, they generalized
the results of Kobayashi, Murakami and Murakami~\cite{kobaiashi} for
the sl(2)-chiral Potts model. The connection of such invariants with the
Seifert matrix has been studied by Goldschmidt and Jones in~\cite{gold}.

Following a similar scheme we study the three-dimensional
integrable {\bbm} from the point of view of the link theory.
We generalize the results of  ref.~\cite{djmm2} to the
R-matrix with spectral parameters associated to the {\bbm}.
We show that choosing
suitable limits of the spectral parameters this matrix gives
cyclic representations of the \bg.
In the trigonometric case we classify all possible spectral limits
which produce {\bg} representations. We prove that
for some of them we get cyclotomic link invariants,  while
for other limits of the
rapidity variables (spectral parameters) the R-matrix of the {\bbm}
gives tangle invariants. Such invariants are generalizations of the
cyclotomic invariants previously mentioned. \footnote{This work
is mainly based on the thesis of B.~L.~Cerchiai,(B.~L.~Cerchiai,
"Modelli di Baxter-Bazhanov e di Potts chirale e teoria dei nodi",academic year
1992-93) in fulfilment of the requirements for the degree (laurea) in Physics.}

\topskip 9mm

\newpage

\newsection{The \hspace{0.5pt} 3-Dimensional \hspace{0.5pt} Baxter-Bazhanov
Model and its 2-Dimensional Reduction}

The {\bbm}~\cite{baxbaz1} is an integrable three-dimensional
IRF(=inter \- action-round-a-face) model.
This means that it is defined on a simple cubic lattice $\call$ and
that a spin variable $\sigma$ is placed at each site of $\call$.
{}From the point of view of statistical mechanics
the {\bbm} depends on two integer parameters $N(N\geq2)$ and n. $N$ is the
number of values that each spin $\sigma$ can take,
while n is one of the lattice dimensions
(=number of elementary cubes in a fixed direction, e.g. in front-to-back
direction).

The elementary cube of $\call$ is shown in the following figure:

\begin{figure}[h]
\vspace{10pt}
\begin{picture}(70,70)(-180,0)
\put(0,0){\framebox(40,40){ }}
\put(0,0){\circle*{5}}
\put(40,0){\circle*{5}}
\put(0,40){\circle*{5}}
\put(40,40){\circle*{5}}
\put(0,0){\line(-1,1){20}}
\put(0,40){\line(-1,1){20}}
\put(40,40){\line(-1,1){20}}
\put(-20,20){\line(0,1){40}}
\put(-20,60){\line(1,0){40}}
\multiput(-20,20)(3,0){13}{\line(1,0){1}}
\multiput(20,20)(0,3){13}{\line(0,1){1}}
\multiput(20,20)(2,-2){10}{\line(1,0){1}}
\put(-20,20){\circle*{5}}
\put(20,20){\circle*{5}}
\put(20,60){\circle*{5}}
\put(-20,60){\circle*{5}}
\put(0,-12){e}
\put(40,-12){d}
\put(0,48){a}
\put(40,48){f}
\put(-20,8){c}
\put(20,8){h}
\put(-20,68){g}
\put(20,68){b}
\end{picture}
\vspace{10pt}
\caption{Elementary cell}
\label{cell}
\end{figure}
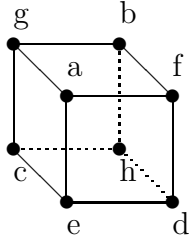

In order to define the Boltzmann weight of the elementary cell
shown in figure~{\ref{cell}} it is necessary to introduce some
notation first.

Let $x$ be a complex parameter and $k,l,m$ three integers,
$0\leq k,l,m \leq N-1$.
Let $\omega$ be a primitive N-th root of unity
\beq
\omega=e^{\frac{2 \pi i}{N}}
\eeq
and
\beq
\omega^{1/2}=e^{\frac{\pi i}{N}}.
\eeq
Let $\Phi$ and $s$ be the functions defined by
\bea
\Phi(x)=(\omega^{1/2})^{N(N+l)} \\
s(k,l)=\omega^{kl}.
\eea
Notice that
\bea
s(k+N,l)=s(k,l+N)=s(k,l) \\
s(k+l,m)=s(k,m)s(l,m) \\
\Phi(k+l)=\Phi(k)\Phi(l)s(k,l)
\eea
Moreover, let $w(x,l)$ be the function defined by
\beq
\frac{w(x,l)}{w(x,0)}=(\Delta(x))^l\prod_{k=1}^l(1-\omega^kx)^{-1}
\label{w}
\eeq
where
\beq
\Delta(x)=(1-x^N)^{1/N}.
\label{delta}
\eeq
With this definition the function $w(x,l)$ is
fixed up to the overall normalization factor $w(x,0)$, while $\Delta$ is
fixed up to the choice of a phase when taking the root.
In particular it is possible to impose the following condition on $w(x,0)$
\beq
w(x,0)=w(\omega^kx,0).
\label{normaw}
\eeq
Applying this condition~(\ref{normaw}) to the definition of $w$~(\ref{w}),
it follows immediately
\beq
w(x,0)w(x,l+k)=w(x,k)w(\omega^kx,l).
\eeq
Having introduced all this notation,
following ref.~\cite{baxbaz1} the Boltzmann weight of the elementary cube
shown in figure~\ref{cell} is constructed as
\beq
W(a \mid e,f,g \mid b,c,d \mid h)=\sum_{\sigma=0}^{N-1}
v_{\sigma}(a \mid e,f,g \mid b,c,d \mid h)
\label{boltz1}
\eeq
with
\bea
\lefteqn{v_{\sigma}(a \mid e,f,g \mid b,c,d \mid h)=} \nonumber \\
& &\frac{w(\frac{p'}{p},e-c-d+h)}{w(\frac{p'}{p},a-g-f+b)}
s(c-h,d-h)s(g,a-g-f+b)
\label{boltz2} \\
& \times & \{
\frac
{w(\frac{p}{q},d-h-\sigma)w(\frac{q'}{p},\sigma-f+b)
w(\frac{p'}{q'},a-g-\sigma)}
{w(\frac{p'}{q},e-c-\sigma) (\Phi(a-g-\sigma))^{-1}} s(\sigma,a-c-f+h) \}.
\nonumber
\eea
The parameters $p,p',q,q'$ are the so-called spectral parameters.
To stress the dependence of $W$ on these parameters,
it would be more correct to write
\[W=W[p,p',q,q'].\]

\begin{figure}[t]
\vspace{10pt}
\begin{picture}(140,140)(-180,-10)
\put(0,0){\dashbox(80,80){ }}
\put(0,0){\circle*{5}}
\put(80,0){\circle*{5}}
\put(0,80){\circle*{5}}
\put(80,80){\circle*{5}}
\put(0,0){\line(-2,1){40}}
\put(80,0){\line(-2,1){40}}
\put(0,80){\line(-2,1){40}}
\put(80,80){\line(-2,1){40}}
\put(-40,20){\dashbox(80,80){ }}
\put(-40,20){\circle*{5}}
\put(40,20){\circle*{5}}
\put(40,100){\circle*{5}}
\put(-40,100){\circle*{5}}
\put(0,-12){e}
\put(80,-12){d}
\put(0,88){a}
\put(80,88){f}
\put(-40,8){c}
\put(40,8){h}
\put(-40,108){g}
\put(40,108){b}
\put(0,0){\line(2,5){40}}
\put(80,0){\line(-6,5){120}}
\put(-40,20){\line(2,1){120}}
\put(40,20){\line(-2,3){40}}
\put(20,50){\circle*{5}}
\put(30,47){$\sigma$}
\end{picture}
\caption{Interactions in the elementary cube}
\label{inter}
\end{figure}
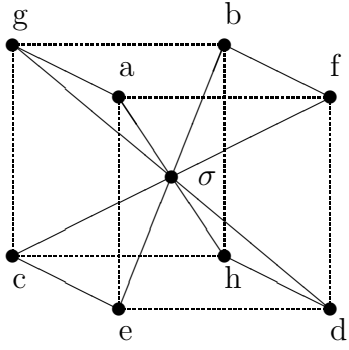

Notice that the spins are seen as elements of $\zbn$ and that
$W$ depends only on the pairwise differences of adiacent spins.
This means that it is consistent to assume the following equivalence
relation between the spins
\beq
\alpha \sim \beta \Longleftrightarrow \alpha_i-\alpha_{i+1}=\beta_i-
\beta_{i+1} \makebox{    $\forall i=1,\ldots,n$}
\label{equivalent}
\eeq
In the expressions~(\ref{boltz1}),~(\ref{boltz2}) $\sigma$ can be interpreted
as a spin at the centre of the cube.
The elementary interactions are shown in figure~\ref{inter}.

This means that in fact we are not considering
a simple cubic lattice, but a body-centred cubic lattice.
Bazhanov and Baxter have noticed~\cite{baxbaz1} that the model
obtained in this way is an Ising type model.
Thus it turns out that (up to an overall normalization factor,
a site-type,edge-type and face-type equivalence transformation)
W satisfies the tetrahedron equation~\cite{baxter,kash1,kash2},
which guarantees that the model is integrable
\bea
\sum_d & W(a_4 \mid c_2,c_1,c_3 \mid b_1,b_3,b_2 \mid d) &
W'(c_1 \mid b_2,a_3,b_1 \mid c_4,d,c_6 \mid b_4) \nonumber \\
\times & W''(b_1 \mid d,c_4,c_3 \mid a_2,b_3,b_4 \mid c_5) &
W'''(d \mid b_2,b_4,b_3 \mid c_5,c_2,c_6 \mid a_1) \nonumber \\
=\sum_d & W'''(b_1 \mid c_1,c_4,c_3 \mid a_2,a_4,a_3 \mid d) &
W''(c_1 \mid b_2,a_3,a_4 \mid d,c_2,c_6 \mid a_1) \nonumber \\
\times & W'(a_4 \mid c_2,d,c_3 \mid a_2,b_3,a_1 \mid c_5) &
W(d \mid a_1,a_3,a_2 \mid c_4,c_5,c_6 \mid b_4).
\label{tetrahedron}
\eea
In this this equation $W=W(P),W'=W(P'),W''=W(P''),W'''=W(P''')$,
where
\bea
P=(x_1,x_2,x_3,x_4), & P'=(x_1',x_2',x_3',x_4'), \nonumber \\
P''=(x_1'',x_2'',x_3'',x_4''), & P'''=(x_1''',x_2''',x_3''',x_4''')
\eea
with $(x_1,x_2,x_3,x_4)=(q,q',p,p')$
and the primes are added to the $x$'s in corrispondence with primes of the
$P$'s. Defining further
\beq
x_{ij}=x_i\Delta(x_j/x_i),
\eeq
the tetrahedron equations~(\ref{tetrahedron}) hold provided
the points $P,P',P'',P'''$ are constrained to satisfy
\[
\begin{array}{llr}
{\displaystyle \frac{x_2}{x_1}=\frac{x_2'}{x_1'},}
& {\displaystyle \frac{x_{12}}{x_1}=\frac{x_{12}'}{x_1'},}
& {\displaystyle \frac{x_3}{\omega x_4}=\frac{x_2'''}{x_1'''},} \\
{\displaystyle \frac{x_{34}}{\omega^{1/2}x_4}=\frac{x_{12}'''}{x_1'''},} &
{\displaystyle \frac{\omega^{1/2}x_{13}x_{24}}{x_{14}x_{32}}=
\frac{x_1''}{x_2''},} &
{\displaystyle \frac{\omega^{1/2}x_{12}x_{34}}{x_{14}x_{32}}=
\frac{x_{12}''}{x_2''},}
\end{array}
\]
\beq
\begin{array}{llr}
{\displaystyle \frac{x_{14}'x_{32}'}{x_{13}'x_{24}'}=
\frac{x_{14}''x_{32}''}{x_{13}''x_{24}''},} &
{\displaystyle \frac{x_{12}'x_{34}'}{x_{13}'x_{24}'}=
\frac{x_{12}''x_{34}''}{x_{13}''x_{24}''},} &
{\displaystyle \frac{x_3''}{x_4''}=\frac{x_3'''}{x_4'''},} \\
{\displaystyle \frac{x_{34}''}{x_4''}=\frac{x_{34}'''}{x_4'''},}
& {\displaystyle \frac{x_4'}{x_3'}=
\frac{x_{13}'''x_{24}'''}{\omega^{1/2} x_{14}'''x_{32}'''},}
 &
{\displaystyle \frac{x_{34}'}{x_3'}
=\frac{x_{12}'''x_{34}'''}{x_{14}'''x_{32}'''},} \\
{\displaystyle
\frac{\omega^{1/2} x_{32}x_4'x_{24}''x_2'''}{x_3x_{24}'x_2''x_{24}'''}=1}
, & & {\displaystyle
\frac{x_{13}x_1'x_{14}''x_1'''}{x_1x_{14}'x_1''x_{14}'''}=1,} \\
{\displaystyle \frac{x_{14}x_4'x_{14}''x_4'''}{x_4x_{14}'x_4''x_{24}'''}=1,}
& & {\displaystyle
\frac{\omega^{1/2}x_{13}x_3'x_{13}''x_2'''}{x_3x_{13}'x_1''x_{32}'''}=1.}
\end{array}
\eeq

At this point it is useful to consider also the Boltzmann weight $S$ of a
parallelepiped~$\calp$ formed by a whole line of n cubes in
front-to-back direction with periodic boundary conditions. Let
\bea
\alpha=(\alpha_1,\ldots,\alpha_n), & \beta=(\beta_1,\ldots,\beta_n),
\nonumber \\
\gamma=(\gamma_1,\ldots,\gamma_n), & \delta=(\delta_1,\ldots,\delta_n)
\eea
denote the spins on the edges of $\calp$.

\begin{figure}[h]
\vspace{10pt}
\begin{picture}(140,140)(-180,-10)
\put(0,0){\framebox(40,40){}}
\multiput(0,0)(-20,20){5}{\circle*{5}}
\multiput(0,40)(-20,20){5}{\circle*{5}}
\multiput(40,40)(-20,20){5}{\circle*{5}}
\put(40,0){\circle*{5}}
\put(0,0){\line(-1,1){45}}
\put(-55,55){\line(-1,1){25}}
\multiput(0,40)(40,0){2}{\line(-1,1){45}}
\multiput(-55,95)(40,0){2}{\line(-1,1){25}}
\put(-17,-5){$\alpha_1$}
\put(48,-5){$\beta_1$}
\put(-17,40){$\delta_1$}
\put(48,40){$\gamma_1$}
\put(-37,15){$\alpha_2$}
\put(-37,60){$\delta_2$}
\put(28,60){$\gamma_2$}
\put(8,80){$\gamma_3$}
\put(-57,80){$\delta_3$}
\put(-57,35){$\alpha_3$}
\put(-77,55){$\alpha_n$}
\put(-77,100){$\delta_n$}
\put(-8,100){$\gamma_n$}
\put(-97,75){$\alpha_1$}
\put(-97,120){$\delta_1$}
\put(-28,120){$\gamma_1$}
\multiput(-20,60)(-20,20){4}{\line(1,0){40}}
\multiput(-20,60)(-20,20){4}{\line(0,-1){40}}
\put(-60,60){\line(1,0){20}}
\put(-20,100){\line(0,-1){20}}
\end{picture}
\caption{Parallelepiped $\calp$}
\label{parallelepiped}
\end{figure}
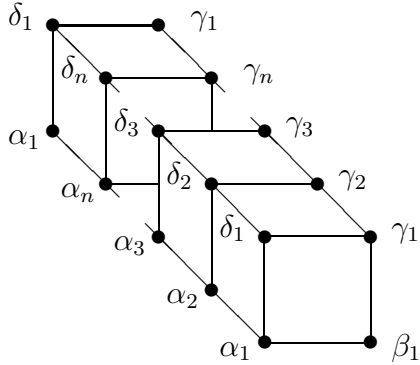

Then
\beq
S(\alpha,\beta,\gamma,\delta)=\prod_{i \epsilon \zsn} W(\delta_i \mid
\alpha_i,\gamma_i,\delta_{i+1} \mid \gamma_{i+1},\alpha_{i+1},\beta_i
\mid \beta_{i+1})
\label{s}
\eeq
Further, following Baxter and Bazhanov, let's introduce also
a slightly modified model.
Let's substitute the variable $\sigma$ with the difference of two new
spins in front-to-back direction.
\beq
\sigma=\mu-\mu'.
\eeq
This means that considering a row of n cubes in front-to-back direction
the following constraint is imposed on the variable $\sigma$
\beq
\sum_{i \epsilon \zsn} \sigma_i=0 \pmod {N}
\eeq
The model obtained with this change of boundary conditions is called
by Baxter and Bazhanov the "modified model".
The Boltzmann weight of the parallalelepiped $\calp$ formed by a line
of cubes in front-to-back direction is denoted $S_0$.
\beq
S_0(\alpha,\beta,\gamma,\delta)=\prod_{i \epsilon \zsn} \sum_{\mu_i=0}^{N-1}
v_{\mu_i-\mu_{i+1}}(\delta_i \mid
\alpha_i,\gamma_i,\delta_{i+1} \mid \gamma_{i+1},\alpha_{i+1},\beta_i
\mid \beta_{i+1})
\label{s01}
\eeq
with
\bea
\lefteqn{v_{\mu_i-\mu_{i+1}}(\delta_i \mid \alpha_i,\gamma_i,\delta_{i+1}
\mid \gamma_{i+1},\alpha_{i+1},\beta_i \mid \beta_{i+1})=} \nonumber \\
& &\frac{w(\frac{p'}{p},\alpha_i-\alpha_{i+1}-\beta_i+\beta_{i+1})}
{w(\frac{p'}{p},\delta_i-\delta_{i+1}-\gamma_i+\gamma_{i+1})}
s(\alpha_{i+1}-\beta_{i+1},\beta_i-\beta_{i+1}) \nonumber \\
& \times & s(\delta_{i+1},\delta_i-\delta_{i+1}-\gamma_i+\gamma_{i+1})
\label{s02} \\
& \times &\{
\frac
{w(\frac{p}{q},\beta_i-\beta_{i+1}-\mu_i+\mu_{i+1})w(\frac{q'}{p},
\mu_i-\mu_{i+1}-\gamma_i+\gamma_{i+1})}
{w(\frac{p'}{q},\alpha_i-\alpha_{i+1}-\mu_i+\mu_{i+1})
(\Phi(\delta_i-\delta_{i+1}-\mu_i+\mu_{i+1}))^{-1}
}
\nonumber \\
& \times &
w(\frac{p'}{q'},\delta_i-\delta_{i+1}-\mu_i+\mu_{i+1})
s(\mu_i-\mu_{i+1},
\delta_i-\alpha_{i+1}-\gamma_i+\beta_{i+1}) \}. \nonumber
\eea

The key idea of ref.~\cite{baxbaz1} is to describe the {\bbm} as an
integrable generalized chiral Potts model~\cite{bkms,djmm1,nakanishi}
in the IRF presentation by the following prescription.

\begin{figure}[h]
\vspace{10pt}
\begin{picture}(140,140)(-175,-10)
\put(10,0){\line(1,0){40}}
\put(30,10){\vector(0,1){40}}
\put(10,0){\circle*{5}}
\put(50,0){\circle*{5}}
\put(-4,0){$\alpha$}
\put(60,0){$\beta$}
\multiput(80,55)(-20,20){5}{\line(1,0){40}}
\multiput(80,55)(40,0){2}{\line(-1,1){45}}
\multiput(27,108)(40,0){2}{\line(-1,1){27}}
\multiput(80,55)(-20,20){5}{\circle*{5}}
\multiput(120,55)(-20,20){5}{\circle*{5}}
\put(60,55){$\alpha_1$}
\put(40,75){$\alpha_2$}
\put(20,95){$\alpha_3$}
\put(0,115){$\alpha_n$}
\put(-20,135){$\alpha_1$}
\put(128,55){$\beta_1$}
\put(108,75){$\beta_2$}
\put(88,95){$\beta_3$}
\put(68,115){$\beta_n$}
\put(48,135){$\beta_1$}
\put(55,-15){$\beta=(\beta_1,\ldots,\beta_n)$}
\put(-47,-15){$\alpha=(\alpha_1,\ldots,\alpha_n)$}
\end{picture}
\caption{Reduction procedure}
\label{reduction}
\end{figure}
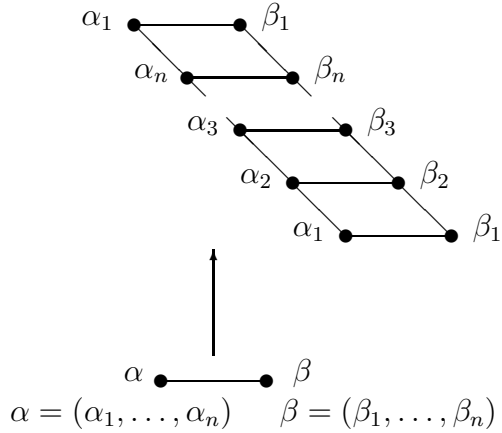

\noindent
For this aim, one starts from an edge of the bidimensional lattice
on which the chiral Potts model is defined. This edge is extended
in a third additional dimension perpendicular to the plane
of the two-dimensional lattice to form a rectangle consisting of
n squares. The two spins $\alpha=(\alpha_1,\ldots,\alpha_n)$
and $\beta=(\beta_1,\ldots,\beta_n)$ located at the vertices of
the two-dimensional lattice are placed on the edges of the rectangle,
as shown in figure~\ref{reduction}.
Cyclic boundary conditions are assumed in the new dimension, considering
the spins $\alpha_1,\beta_1$ as next to $\alpha_n,\beta_n$ respectively.
Doing this construction for all edges of the two-dimensional lattice,
it becomes the three-dimensional cubic lattice $\call$ with N-valued
spins at each site.

Baxter and Bazhanov have proved that the weight function
$W_{pq}(\alpha,\beta)$ of the chiral Potts
model associated to an edge can be written in the following form
\beq
\frac{W_{pq}(\alpha,\beta)}{W_{pq}(0,0)}=
\prod_{i=1}^n\{\omega^{(\beta_i-\beta_{i+1})(\alpha_{i+1}-\beta_{i+1})}
w(\frac{p_i}{q_i},\alpha_i-\alpha_{i+1}-\beta_i+\beta_{i+1})\},
\label{edge}
\eeq
Notice that the rapidity variables in~(\ref{edge}) form a n-vector
\[p=(p_1,\ldots,p_n),q=(q_1,\ldots,q_n)\]
exactly as the spins do.
In the three-dimensional interpretation the weight $W_{pq}$
is associated to the whole rectangle constructed in figure~\ref{reduction}.
This three-dimensional re-interpretation of the two-dimensional
statistical model is allowed by the factorization property~(\ref{edge})
of the Boltzmann weight:the i-th term in the product depends only on
the four spins $\alpha_i,\beta_i,\alpha_{i+1},\beta_{i+1}$
located at the vertices of the i-th elementary square in the rectangle.
Notice that not all two-dimensional integrable models have this
factorization property.

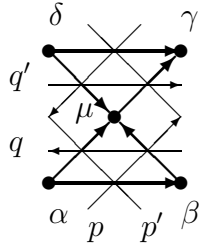
\begin{figure}[h]
\vspace{10pt}
\begin{picture}(70,80)(-180,-20)
\thicklines
\multiput(0,0)(0,50){2}{\vector(1,0){50}}
\multiput(0,0)(50,0){2}{\circle*{5}}
\multiput(0,50)(50,0){2}{\circle*{5}}
\put(25,25){\circle*{5}}
\put(10,25){$\mu$}
\put(0,60){$\delta$}
\put(50,60){$\gamma$}
\put(0,-15){$\alpha$}
\put(50,-15){$\beta$}
\put(-15,37){$q'$}
\put(-15,12){$q$}
\put(15,-20){$p$}
\put(35,-20){$p'$}
\multiput(0,0)(25,25){2}{\vector(1,1){23}}
\put(50,0){\vector(-1,1){23}}
\put(0,50){\vector(1,-1){23}}
\thinlines
\put(0,37){\vector(1,0){50}}
\put(50,12){\vector(-1,0){50}}
\put(15,-10){\vector(1,1){35}}
\put(50,25){\line(-1,1){35}}
\put(35,60){\vector(-1,-1){35}}
\put(0,25){\line(1,-1){35}}
\end{picture}
\caption{Elementary star}
\label{star}
\end{figure}

Let us now consider the star of figure~\ref{star}.

\noindent
Corresponding to this configuration we define the star-Boltzmann
weight $W_{star}^{(1)}$ of the IRF chiral Potts model
\bea
\lefteqn{W_{star}^{(1)}(p,p',q,q'\mid \alpha,\beta,\gamma,\delta)=}
\nonumber\\
& &\sum_{\mu}\frac{W_{p'p}(\alpha,\beta)}{W_{p'p}(\delta,\gamma)}
\frac{
W_{pq}(\beta,\mu)W_{q'p}(\mu,\gamma)W_{p'q'}(\delta,\mu)}{W_{p'q}(\alpha,\mu)
}
\label{potts}
\eea
whose $W_{ij},(i,j=p,p',q,q')$ are the edge-Boltzmann weights defined
in~(\ref{edge}).
It turns out that $W_{star}^{(1)}$ satisfy the Yang-Baxter
equation~\cite{djmm1,nakanishi}
\bea
\sum_{\sigma}W_{star}^{(1)}(p,p',q,q' \mid \alpha,\beta,\gamma,\sigma) &
W_{star}^{(1)}(p,p',r,r'\mid \sigma,\gamma,\delta,\epsilon) \nonumber \\
W_{star}^{(1)}(q,q',r,r'\mid \alpha,\sigma,\epsilon,\kappa)= &
{\displaystyle \sum_{\sigma}W_{star}^{(1)}(q,q',r,r'
\mid \beta,\gamma,\delta,\sigma)}
\nonumber \\
W_{star}^{(1)}(p,p',r,r' \mid \alpha,\beta,\sigma,\kappa) &
W_{star}^{(1)}(p,p',q,q' \mid \kappa,\sigma,\delta,\epsilon).
\eea

The connection between the chiral Potts model and the {\bbm} arises,
because it turns out that the Boltzmann weight of the row of
cubes in front-to-back direction $\calp$ in the modified model
exactly coincides with $W_{star}^{(1)}$.
\beq
S_0(\alpha,\beta,\gamma,\delta)=W_{star}^{(1)}(\alpha,\beta,\gamma,\delta).
\label{sw}
\eeq

Then in order to construct (cyclic) representations of the braid group,
the usual procedure~\cite{djmm1,djmm2} is to map by a Wu-Kadanoff-Wegener
like tranformation the IRF R-matrix defined by $W_{star}^{(1)}$ to a
vertex-type one, and hence to show that it is an intertwiner
of the (cyclic) representations of the quantum group
$U_q(\hat{gl}_n)$.
The main result of this paper is to show in the next sections that
by choosing some suitable limits of the
spectral parameters characterizing the IRF-R-matrix~(\ref{potts}) of the
three-dimensional Baxter-Bazhanov model, one may obtain directly
cyclic representations of the \bg, similarly to the two-dimensional
case~\cite{aku1,aku2}.

\newsection{The Cyclotomic Invariants}

In order to construct cyclic representations of the {\bg} and the related
cyclotomic invariants, the starting point is the construction of a
$\complex^*$-algebra $\cala(c)$ and of a functor $\calf$ from
the category of the uniform oriented tangles $\calt_M^M$ to
$\cala(c)$~\cite{djmm2}.
Let us introduce some notations first.
Let $L$ be a free $\zbn$-module of rank $n-1$ and suppose it is given by
the exact sequence
\[0 \rightarrow \zbn {\rm Ker } \pi=\zbn (1,\ldots,1) \rightarrow \zbn^n
\rightarrow L \rightarrow 0 \]
This means that it is possible to write the elements of $L$ as
\[ \alpha=(\alpha_1,\ldots,\alpha_n) \]
with the equivalence relation~(\ref{equivalent}), which
implements the $\zbn^{n-1}$ symmetry
of the \bbm.
Next let's introduce the non-singular bilinear form $B$ on $L$
\beq
B(\alpha,\beta)=-\sum_{i \epsilon \zsn} \alpha_i(\beta_i-\beta_{i+1})
\eeq
which corresponds to the $n \times n$ matrix
\beq
B_{ij}= \left\{ {\begin{array}{lll}
-1 & \mbox{if $i=j$} \\
1 & \mbox{if $i=j-1 \pmod{n}$} \\
0 & \mbox{otherwise}
\end{array}}
\right.
\eeq
Let $A(\alpha,\beta)$ be twice the skew-symmetric part of $B(\alpha,\beta)$
\beq
A(\alpha,\beta)=B(\alpha,\beta)-B(\beta,\alpha)
\eeq
These definitions are consistent, since B respects the
equivalence relation~(\ref{equivalent}).
Next we assigne a configuration $c$. By this we mean that we give
a map \linebreak $c:~\{1,\ldots,M\}~\mapsto~\{ \pm 1 \}$
that we write $c=(c(1),\ldots,c(M))$.
We interpret $c$ as an object of the category $\calt_M^M$ of the uniform
oriented tangles~\cite{yetter1,yetter2}.

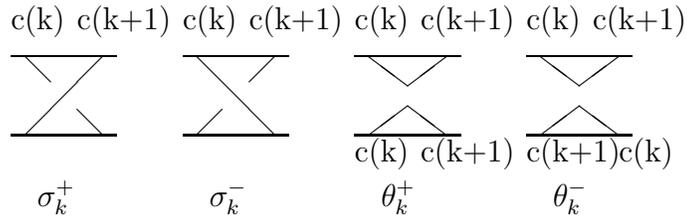
\begin{figure}[h]
\vspace{10pt}
\begin{picture}(300,70)(-100,-30)
\multiput(0,0)(65,0){4}{\line(1,0){40}}
\multiput(0,30)(65,0){4}{\line(1,0){40}}
\put(5,0){\line(1,1){30}}
\multiput(35,0)(-20,20){2}{\line(-1,1){10}}
\multiput(70,0)(20,20){2}{\line(1,1){10}}
\put(100,0){\line(-1,1){30}}
\multiput(135,0)(65,0){2}{\line(4,3){15}}
\multiput(165,0)(65,0){2}{\line(-4,3){15}}
\multiput(135,30)(65,0){2}{\line(4,-3){15}}
\multiput(165,30)(65,0){2}{\line(-4,-3){15}}
\multiput(0,40)(65,0){4}{c(k)}
\multiput(25,40)(65,0){4}{c(k+1)}
\multiput(130,-10)(100,0){2}{c(k)}
\multiput(155,-10)(40,0){2}{c(k+1)}
\put(10,-27){$\sigma_k^+$}
\put(75,-27){$\sigma_k^-$}
\put(140,-27){$\theta_k^+$}
\put(205,-27){$\theta_k^-$}
\end{picture}
\caption{Elementary tangles}
\label{tangles}
\end{figure}

The algebra $\cala(c)$ is a $\complex$-algebra with a unit $1$.
If $M=0$ or $M=1$, $\cala(c)$ is simply $\complex$. If $M\geq2$,
$\cala(c)$ is the algebra with generators $x_k^{\alpha}=x_k^{\alpha}(c)$
where $1\leq k\leq M-1$ and $\alpha \epsilon L$. We impose the relations
\beq
\begin{array}{lllllll}
x_k^0 &=1,& \\
x_k^{\alpha}x_k^{\beta} & =\omega^{A(\beta,\alpha)/2}x_k^{\alpha+\beta}
& {\mbox{if $(c(k),c(k+1))=(1,1)$}} \\
 &=\omega^{A(\alpha,\beta)/2}x_k^{\alpha+\beta}
& {\mbox{if $(c(k),c(k+1))=(-1,-1)$}} \\
 &=x_k^{\alpha+\beta}
& {\mbox{if $(c(k) \not =c(k+1)$}} \\
x_k^{\alpha}x_{k+1}^{\beta} & =\omega^{B(\beta,\alpha)}x_{k+1}^{\beta}
x_k^{\alpha}
& {\mbox{if $c(k+1)=-1$}} \\
 &=\omega^{B(\alpha,\beta)}x_{k+1}^{\beta}x_k^{\alpha}
& {\mbox{if $c(k+1)=1$}} \\
x_k^{\alpha}x_{k'}^{\beta} & =x_{k'}^{\beta}x_{k}^{\alpha}
& {\mbox{if $|k-k'|\geq2$}}
\end{array}
\label{algebra}
\eeq
On the algebra $\cala(c)$ there is a linear involution, which is defined
by its action on the generators
\beq
(x_k^{\alpha})^*=x_k^{-\alpha}.
\eeq
In terms of the operators $x_k^{\alpha}$ it is possible to define
the operators describing the images of the functor $\calf$ of the elementary
tangles as
\beq
\begin{array}{lll}
T_k(c) & =\frac{1}{\sqrt D}\sum_{\alpha \epsilon L} \omega^{
-B(\alpha,\alpha)/2}x_k^{\alpha}(c)
& {\makebox{if $c(k)=c(k+1)$}} \\
 &=\frac{1}{D}\sum_{\alpha,\beta \epsilon L}\omega^{B(\beta,\beta)+
B(\alpha,\beta)} x_k^{\alpha}(c)
& {\makebox{if $c(k) \not = c(k+1)$}} \\
E_k(c) &=\frac{1}{\sqrt D} \sum_{\alpha \epsilon L}
 x_k^{\alpha}(c)
& {\makebox{if $c(k) \not = c(k+1)$}} \\
\end{array}
\label{operators}
\eeq
where $D=N^{n-1}$
and $E_k$ is defined only when $c(k) \not = c(k+1)$.
Remind that the functor $\calf$ from the category $\calt_M^M$ of the uniform
oriented tangles is constructed as follows~\cite{djmm2}.
First the morphisms of $\calt_M^M$ are generated by the
elementary tangles shown in figure~\ref{tangles}.

\noindent
As a consequence of the defining relations~(\ref{operators}) of the morphisms
$T_k,E_k$ and using the commutation relations~(\ref{algebra}) of the
$x_k^{\alpha}$,
it is possible to verify that the elements~(\ref{operators})
satisfy the following relations~(\ref{frel}), in which the strings
should be oriented in all possible ways
\bea
T_k^* =T_k^{-1} \nonumber \\
E_k^* =E_k \nonumber \\
T_k T_{k'}=T_{k'} T_k {\makebox{  if $|k-k'|\geq2$}} \nonumber \\
E_k T_{k'}=T_{k'} E_k {\makebox{  if $|k-k'|\geq2$}} \nonumber \\
E_k^2=E_k
\label{frel} \\
E_k E_{k+1} E_k =E_k \nonumber \\
T_k T_{k+1} T_k=T_{k+1} T_k T_{k+1} \nonumber \\
T_k E_{k+1} E_k =T_{k+1}^{-1} E_k \nonumber \\
T_k^{2N}=(scalar) 1. \nonumber
\eea
Notice that the first relation means that $T_k$ is unitary,
while the third and the seventh are the {\bg} relations.
The (\ref{frel}) are some of the defining relations of the class
of the morphisms $h(c,c)$ in the category $\calt_M^M$.
Date, Jimbo, Miki and Miwa~\cite{djmm2} have
shown that the other relations defining this class are
satisfied too.
This means that the functor $\calf$ mapping $\calt_M^M$ to $\cala(c)$
defined by
\beq
\begin{array}{ll}
\calf(c)=\cala(c) \\
\calf(\sigma_k^{\pm})=\calm(T_k(c)^{\pm 1}), &
\calf(\theta_k^{\pm})=\calm(E_k(c)^{\pm 1}).
\end{array}
\eeq
is well-defined.
Here $\calm(a)\epsilon End(\cala(c))$ denotes the left multiplication
by $a \epsilon \cala$.
Moreover, let $\tr$ denote the usual matrix trace on $End(\cala(c))$
normalized as $\tr(1)=1$.
Notice that if $c(k)=1$, $\forall k$, or $c(k)=-1$, $\forall k$, then the
tangles $T_k$ give a representation of the ordinary \bg. In that case it
is possible to consider also the right multiplication by elements of
$\cala(c)$ and we obtain a right-regular representation of the \bg, not only
a left-regular one.
Then if n=2 and
$c(k)=1$,$\forall k$, or $c(k)=-1$,$\forall k$,
one finds the "generalized skein relations"
\beq
T_k^l=\sum_{i=0}^{l-1} A_i T_k^i
\label{skein}
\eeq
In this equation the order of the skein relation is given by
\beq
l=\left\{\begin{array}{ll}
\frac{N+2}{2} &{\makebox{if $N$ even}} \\
\frac{N+1}{2} &{\makebox{if $N$ odd}}
\end{array}
\right.
\label{order}
\eeq
Formula~(\ref{order}) is valid when the equation
\beq
x^2=1 \pmod{N}
\eeq
has only the two solutions $x=1 \pmod{N}$ and $x=N-1 \pmod{N}$.
In particular it is valid for the prime numbers.
It shows that for $n=2,N=2,3$ the algebra defined by the operators
$T_k$ can be expressed in terms of the Hecke algebra~\cite{aku1,aku2}.
More generally operators satisfying generalized skein relations
like~(\ref{skein}) can be obtained with a "cabling procedure"
starting from the generators of the Hecke algebra.
In equation ~(\ref{skein}) the coefficients $A_i$ are the solutions
of the linear system
\beq
\begin{array}{l}
\frac{1}{D^{l/2}} \sum_{\alpha(2),\ldots,\alpha(l) \epsilon L}
\prod_{i=1}^{l-1} \omega^{B(\alpha(i+1),\alpha(i)-\alpha(i+1))}= \\
\delta_{\alpha(1),(0,0)} A_0 +\sum_{r=1}^{l-1}\frac{A_r}{D^{r/2}}
\sum_{\alpha(2),\ldots,\alpha(r) \epsilon L}
\prod_{i=1}^{r-1} \omega^{B(\alpha(i+1),\alpha(i)-\alpha(i+1))}
\end{array}
\eeq
With all this preliminaries an invariant of oriented links can
be constructed as follows
\beq
\tau(\hat{t})=D^{(M-1)/2}\tr(\calm(T_{k_1}^{\epsilon(1)} \cdots
T_{k_m}^{\epsilon(m)}))
\makebox{   for $t \epsilon hom(c,c)$}
\label{invariant}
\eeq
if $\hat{t}$ denotes the link obtained by closing the tangle t
and
\beq
\calf(t)=\calm(T_{k_1}^{\epsilon(1)} \cdots T_{k_m}^{\epsilon(m)}).
\label{link}
\eeq
In equation ~(\ref{link}) $\epsilon(i)$ denotes the orientation
of the i-th generator $T_{k_i}$ which appears in the expression of $\calf(t)$.
Notice that the $\tr$ can be defined by its action on the generators
\beq
\tr(x_k^{\alpha_1} \ldots x_k^{\alpha_n})=\left\{
\begin{array}{ll}
1 & {\makebox{if $\alpha_1= \cdots \alpha_n=0$}} \\
0 & {\makebox{otherwise}}
\end{array}
\right.
\label{trace}
\eeq
The quantity~(\ref{invariant}) is invariant under the Markov moves
and gives a cyclotomic invariant of the tangle.
Here by cyclotomic invariant we mean a link invariant defined
through a cyclic representation of the \bg.
In the case that the tangles are associated to braids,
Date et al. have shown that if $\hat{b}$ is a closure of a braid
$b \epsilon B_M$ with $v$ crossings, then~(\ref{invariant}) becomes
\beq
\begin{array}{l}
\tau(\hat{b})=N^{n(M-v-1)/2} \sum_{\alpha \epsilon \zbn^{v-M+1} \otimes
\zbn^n} \omega^{Q(\alpha,\alpha)/2}, \\
{\makebox{    with $b \epsilon B_M$}}.
\end{array}
\eeq
Here $Q(\alpha,\alpha)$ is the bilinear form determined by the matrix
\beq
Q=S \otimes B+S^T \otimes B^T
\label{q}
\eeq
where $S$ is a $(v-M+1) \times (v-M+1)$ Seifert matrix for $\hat{b}$
and T denotes the transposition of a matrix.

Now (\ref{q}) has a topological meaning, since Q
is a presentation matrix for
the $\zet$-module $H_1(M_n,\zbn)$.
Here $M_n$ is the n-th cyclic covering of $S^3$ branched along the link
$\hat{b}$. This means that for N an odd prime number
the module of $\tau(\hat{b})$ can be written as
\beq
|\tau(\hat{b})|=N^{\beta_n/2}
\eeq
where $\beta_n$
is the first Betti number of $M_n$ relatively to the homology group \linebreak
$H_1(M_n,\zbn)$.
Hence, if the quadratic form B is non-singular $\tau$ can be expressed
in terms of products of Alexander-Conway polynomials
associated to the link $\hat{b}$.

\newsection{The Spectral Limits of the IRF-2D-Reduced {\bbmm}
\hspace{0.5pt} R-Matrix \hspace{0.5pt} and the Tangle Invariants}

In this section we shall show that it is possible to obtain directly the
cyclotomic invariants from the Boltzmann weights $S$ of the 3D-{\bbm}
(see section 2), after taking some suitable limits of the
spectral parameters $(p,p',q,q')$. Furthermore, we shall show that
taking other limits of the spectral parameters
it is possible to obtain generalizations of the cyclotomic invariants
from the Boltzmann weights $S_0$ of the modified {\bbm}.
The first step is to define the Yang-Baxter operators
\beq
\begin{array}{l}
(Y_{k0}(p,p',q,q'))_{\alpha(1)\ldots\alpha(M-1)}^{\alpha'(1)\ldots\alpha'(M-1)}
= \\
\frac{1}{D}(\prod_{l \not = k}\delta_{\alpha(l)\alpha'(l)})
S_0(\alpha(k-1),\alpha(k),\alpha(k+1),\alpha'(k))  \\
(Y_k(p,p',q,q'))_{\alpha(1)\ldots\alpha(M-1)}^{\alpha'(1)\ldots\alpha'(M-1)}=
\\
\frac{1}{D}(\prod_{l \not = k}\delta_{\alpha(l)\alpha'(l)})
S(\alpha(k-1),\alpha(k),\alpha(k+1),\alpha'(k))
\end{array}
\eeq
These operators act on a subspace $(\calw^{(0)})^{\otimes M-1}
\subset \calw^{\otimes M-1}$ where $\calw=(C^N)^{\otimes n}$ and
$\calw^{(0)}$ is the subspace generated by the elements of $\calw$
\beq
\xi_{\alpha}=\sum_k \varepsilon_{\alpha_1+k} \otimes \cdots \otimes
\varepsilon_{\alpha_n+k}
\eeq
if {$\varepsilon_i$} is the canonical base of $C^N$ and
$\alpha \epsilon L$.
The subspace $\calw^{(0)}$ has dimension $D=N^{n-1}$, while $\calw$
has dimension $N^n$. But this restriction is necessary in order to implement
the $\zbn^{n-1}$-symmetry of the {\bbm} and hence the equivalence
relation~(\ref{equivalent}).

The Yang-Baxter operators depend on the spectral parameters $(p,p',q,q')$.
In analogy with the standard procedure established by, e.g., Akutsu, Deguchi,
Wadati~\cite{aku1,aku2}, the operators $Y_k$ and $Y_{k0}$ give a
matrix representation of the {\bg} $B_M$ if some
spectral limits on $(p,p',q,q')$ are taken. It turns out that in these limits
$Y_k$ goes either to the left-regular or to the right-regular
representation of the operators $T_k(c)^{\pm 1}$ with $c(k)=1$,$\forall k=1,
\ldots,M$ or $c(k)=-1$, $\forall k=1,\ldots,M$.
To find {\bg} representations the first thing is to look for the
values of the spectral parameters where the model is critical.
This means that we must consider the trigonometric limit,in which
all the elementary cubes in the parallelepiped $\calp$ considered in
section~2 have the same spectral parameters.
This assumption guarantees that the model is homogeneous.
Then we have found the following limits in which we obtain the
left-regular representation of the operators $T_k^{\pm 1},k=1,\ldots,M-1$:
\beq
\begin{array}{llll}
Ia) &p \ll q \ll p'=q'&: Y_k \mapsto T_k(c), &\makebox{  $c(k)=1$,
$\forall k=1,\ldots,M$} \\
Ib) &q \ll p \ll p'=q'&:Y_k \mapsto T_k^{-1}(c), &\makebox{  $c(k)=1$,
$\forall k=1,\ldots,M$} \\
IIa)& p' \ll q' \ll p=q&: Y_k \mapsto T_k(c), &\makebox{  $c(k)=-1$,
$ \forall k=1,\ldots,M$} \\
IIb)& q' \ll p' \ll p=q&: Y_k \mapsto T_k^{-1}(c), &\makebox{  $c(k)=-1$,
$\forall k=1,\ldots,M$}
\end{array}
\eeq

To see this, let's choose the following base of the algebra $\cala(c)$
\beq
\{y(c)=\omega^{\frac{-1}{2}\sum_{k=1}^{M-2}B^{c(k+1)}(\alpha(k),\alpha(k+1))}
x_1^{\alpha(1)}\cdots x_{M-1}^{\alpha(M-1)} \}_{\alpha(i)\epsilon L},
\eeq
where
\beq
B^c(\alpha,\beta)=\left\{\begin{array}{ll}
B(\alpha,\beta) & \makebox{   if $c=1$} \\
B(\beta,\alpha) & \makebox{   if $c=-1$}
\end{array}
\right.
\eeq
The map
\beq
\rho:A(c) \mapsto (\calw^{(0)})^{\otimes M-1}
\eeq
defined by
\beq
\rho(y(c))=\xi_{\alpha(1)} \otimes \cdots \otimes \xi_{\alpha(M-1)}
\eeq
is an isomorphism of \complex$^*$-algebras.
Let's prove that~Ia) is right.
The matrix elements of the operators $\rho T_k \rho^{-1}$
in the case $c(k)=1$, $\forall k$, omitting $\rho$ can be written as
\beq
\begin{array}{lll}
(T_k)_{\alpha(1)\ldots\alpha(M-1)}^{\alpha'(1)\ldots\alpha'(M-1)} \sim
(\prod_{l \not = k} \delta_{\alpha(l)\alpha'(l)}) \omega^{B(\alpha'(k),
\alpha(k))}\\
\times \frac{1}{\sqrt D} \omega^{[1/2 (B(\alpha(k)-\alpha'(k),\alpha(k+1))
-B(\alpha(k-1),\alpha(k)-\alpha'(k))-B(\alpha(k),\alpha(k))
-B(\alpha'(k),\alpha'(k)))]}
\end{array}
\label{opel}
\eeq
where $T_k \sim (something)$ means $\rho T_k \rho^{-1}=(something)$.
The Yang-Baxter operator $Y_k$ in the limit~Ia) gives the same matrix
operators, provided a similarity transformation is made.
To obtain this result, let's calculate the limits of the function
$w$ defined in~(\ref{w}). It results
\beq
\frac{w(x,l)}{w(x,0)}=\left\{\begin{array}{lll}
\Phi(l)^{-1} & \makebox{   if $x \rightarrow \infty$} \\
\delta_{l0} & \makebox{   if $x \rightarrow 1$} \\
1 & \makebox{   if $x \rightarrow 0$}
\end{array}
\right.
\label{limw}
\eeq
Using these limits it is possible to show that
\beq
\frac{W_{pq}(\alpha,\beta)}{W_{pq}(0,0)}=\left\{\begin{array}{lll}
\omega^{B(\alpha,\alpha-\beta)} & \makebox{   if $p/q \rightarrow \infty$} \\
\delta_{\alpha,\beta} & \makebox{   if $p/q \rightarrow 1$} \\
\omega^{B(\beta,\alpha-\beta)} & \makebox{   if $p/q \rightarrow 0$}
\end{array}
\right.
\label{limb}
\eeq
{}From~(\ref{sw}) it follows immediately that in the limit~Ia)
\beq
S(\alpha,\beta,\gamma,\delta)=S_0(\alpha,\beta,\gamma,\delta)=
\omega^{B(\delta,\beta)-B(\delta,\delta)+B(\alpha,\delta-\beta)}
\label{ops}
\eeq
To obtain~(\ref{opel}) from (\ref{ops}) we multiply $S$ by the factor
\beq
\sqrt D
\omega^{\frac{1}{2} [B(\delta,\delta)-B(\beta,\beta)+B(\beta-\delta,\gamma)
+B(\alpha,\beta-\delta)]}.
\label{factor}
\eeq
It is a site-type,edge-type,face-type equivalence transformation and
does not change the factorization
properties nor the partition function of the model.
With the same tools it is possible to see that Ib,IIa,IIb) holds, provided
that $S$ is multiplied by the factor~(\ref{factor}) in the case Ib), and
by the factor
\beq
\sqrt D
\omega^{\frac{1}{2} [B(\beta,\beta)-B(\delta,\delta)-B(\beta-\delta,\alpha)
-B(\gamma,\beta-\delta)]}
\eeq
in the cases IIa) and IIb).
Further, by the same arguments, it is possible to prove that there are other
limits giving the $T_k^{\pm 1}$ in the right-regular representation, obtained
from the left-regular one by transposing the matrices.
These limits are given by
\beq
\begin{array}{llll}
IIIa) &p=q \ll p' \ll q'&: Y_k \mapsto T_k(c), &\makebox{  $c(k)=1$,
$\forall k=1,\ldots,M$} \\
IIIb) &p=q \ll q' \ll p'&:Y_k \mapsto T_k^{-1}(c), &\makebox{  $c(k)=1$,
$\forall k=1,\ldots,M$} \\
IVa)& p'=q' \ll p \ll q&: Y_k \mapsto T_k(c), &\makebox{  $c(k)=-1$,
$\forall k=1,\ldots,M$} \\
IVb)& p'=q' \ll q \ll p&: Y_k \mapsto T_k^{-1}(c), &\makebox{  $c(k)=-1$,
$\forall k=1,\ldots,M$} \end{array}
\eeq
At this point a question arises: is it possible to get other kinds of {\bg}
representations and hence other link invariants starting from the
Yang-Baxter equation of the \bbm?
We fix the configuration of 2M-1 strings
where $c(2k-1)=-1,c(2k)=1$, $\forall k=1,\ldots,M-2$,$c(2M-1)=-1$.
We obtain the following picture for $k=1,\ldots,M-3$
\beq
\begin{array}{llllllllllllllllllllllll}
Va) &p' \ll q' \ll p \ll q: Y_{k0} \mapsto \\
& T_{2k+1}(s_{2k+1}(c))T_{2k}(s_{2k+1}(c))T_{2k+2}(s_{2k+1}(c))
T_{2k+1}(c)^{-1} \\
Vb) &q' \ll p' \ll q \ll p:Y_{k0} \mapsto \\
& T_{2k+1}(s_{2k+1}(c))T_{2k}(s_{2k+1}(c))^{-1}T_{2k+2}(s_{2k+1}(c))^{-1}
T_{2k+1}(c)^{-1} \\
VIa)& p \ll q \ll p' \ll q': Y_{k0} \mapsto \\
& T_{2k+1}(s_{2k+1}(c))^{-1}T_{2k}(s_{2k+1}(c))T_{2k+2}(s_{2k+1}(c))
T_{2k+1}(c) \\
VIb)& q \ll p \ll q' \ll p': Y_{k0} \mapsto \\
& T_{2k+1}(s_{2k+1}(c))^{-1}T_{2k}(s_{2k+1}(c))^{-1}T_{2k+2}(s_{2k+1}(c))^{-1}
T_{2k+1}(c) \\
VIIa) &p \ll q \ll q' \ll p': Y_{k0} \mapsto \\
& T_{2k+1}(s_{2k+1}(c))^{-1}T_{2k}(s_{2k+1}(c))T_{2k+2}(s_{2k+1}(c))^{-1}
T_{2k+1}(c) \\
VIIb) &q \ll p \ll p' \ll q':Y_{k0} \mapsto \\
& T_{2k+1}(s_{2k+1}(c))^{-1}T_{2k}(s_{2k+1}(c))^{-1}T_{2k+2}(s_{2k+1}(c))
T_{2k+1}(c) \\
VIIIa)& q' \ll p' \ll p \ll q: Y_{k0} \mapsto \\
& T_{2k+1}(s_{2k+1}(c))T_{2k}(s_{2k+1}(c))T_{2k+2}(s_{2k+1}(c))^{-1}
T_{2k+1}(c)^{-1} \\
VIIIb)& p' {\ll} q' {\ll} q {\ll} p: Y_{k0} \mapsto \\
& T_{2k+1}(s_{2k+1}(c))T_{2k}(s_{2k+1}(c))^{-1}T_{2k+2}(s_{2k+1}(c))
T_{2k+1}(c)^{-1} \\
IXa) &p \ll p' \ll q \ll q': Y_{k0} \mapsto \\
& T_{2k+1}(s_{2k+1}(c))T_{2k}(s_{2k+1}(c))T_{2k+2}(s_{2k+1}(c))
T_{2k+1}(c) \\
IXb) &q \ll q' \ll p \ll p':Y_{k0} \mapsto \\
& T_{2k+1}(s_{2k+1}(c))^{-1}T_{2k}(s_{2k+1}(c))^{-1}T_{2k+2}(s_{2k+1}(c))^{-1}
T_{2k+1}(c)^{-1} \\
Xa)& p' \ll p \ll q' \ll q: Y_{k0} \mapsto \\
& T_{2k+1}(s_{2k+1}(c))T_{2k}(s_{2k+1}(c))T_{2k+2}(s_{2k+1}(c))
T_{2k+1}(c) \\
Xb)& q' \ll q \ll p' \ll p: Y_{k0} \mapsto  \\
& T_{2k+1}(s_{2k+1}(c))^{-1}T_{2k}(s_{2k+1}(c))^{-1}T_{2k+2}(s_{2k+1}(c))^{-1}
T_{2k+1}(c)^{-1}
\label{lim}
\end{array}
\eeq
Now we must explain the meaning of the products
\[
T_{\pm 2k+1}(s_{2k+1}(c))T_{\pm 2k}(s_{2k+1}(c))
T_{\pm 2k+2}(s_{2k+1}(c))T_{\pm 2k+1}(c)
\]
where $T_{\pm 1 k}=T_k^{\pm 1}$.
We construct a representation $\calr$ of A(c) on
$(\calw^{(0)})^{\otimes {M-1}}$ for the configuration
$c(2k-1)=-1,c(2k)=1$ for $1 \leq k \leq M-2$, $c(2M-1)=-1$ in the following
way.
Notice that adiacent strings have the opposite directions.
Following reference~\cite{djmm2}, we introduce the following operators
acting on $\calw^{(0)}$
\bea
Z_i=1 \otimes \cdots \otimes Z \otimes \cdots \otimes 1
\label{z} \\
X_i=1 \otimes \cdots \otimes X \otimes \cdots \otimes 1
\label{x}
\eea
where $X$ and $Z$ act on the i-th factor $\complex^N$ in $\calw^{(0)}$
and are defined by
\bea
Z_{kl}=\delta_{k,l+1} \\
X_{kl}=\omega^k \delta_{k,l}.
\eea
for $k,l \epsilon Z_N$.
Moreover, using the operators~(\ref{z}),(\ref{x}) we define for
$\alpha \epsilon L$
\bea
X^{\alpha}=X_1^{\alpha_1} \cdots X_n^{\alpha_n} \\
Z^{\alpha}=Z_1^{\alpha_1} \cdots Z_n^{\alpha_n}
\eea
Then the representation $\calr$ is given by
\bea
\calr(x_{2k-1}^{\alpha})=1 \otimes \cdots \otimes Z^{\alpha} \otimes \cdots
\otimes 1
\label{z1} \\
\calr(x_{2k}^{\alpha})=1 \otimes \cdots \otimes X^{B \alpha} \otimes
X^{-B \alpha} \otimes \cdots \otimes 1
\label{z2}
\eea
where the action of $Z^{\alpha}$ in~(\ref{z1}) is on the k-th space,
while the action of $X^{B \alpha} \otimes X^{-B \alpha}$ in~(\ref{z2})
is on the k-th and k+1-th space.
Notice that it is possible to multiply the operators
$T_{k+1}(s_k(c))T_k(c)$ relative to
configurations which differ by a permutation, because the algebras $\cala(c)$
arising from configurations of the same tipe are
canonically isomorphic (see~\cite{djmm2}).
Thus, as a consequence, the matrix elements of the Yang-Baxter operators
$Y_{k0}$ in the limits~V)-X) are exactly the matrix elements
of the products
$T_{\pm 2k+1}(s_{2k+1}(c))T_{\pm 2k}(s_{2k+1}(c))
T_{\pm 2k+2}(s_{2k+1}(c))T_{\pm 2k+1}(c)$
in the representation $\calr$, where in~(\ref{lim}) we have
omitted to write the label $\calr$.

Moreover, we have verified that the trace on the {\bg} representation
given by the operators $Y_{k0}$ in the
limits~V) and VI) enjoies the Markov properties.
This can be verified immediately, by observing that
in the representation $\calr$ the trace has the properties~(\ref{trace}).
Let's show, e.g., the invariance under the Markov move 2 in the case Va).
We define
\beq
\begin{array}{l}
\pi: B_{M-2} \rightarrow (\calw^{(0)})^{M-1} \\
\pi(b_k)=Y_{k0},k=1,\ldots,M-3
\end{array}
\eeq
where the $b_k$ are the {\bg} generators satisfying
the relations
\beq
\begin{array}{ll}
b_k b_{k'}=b_{k'} b_k & \makebox{      for $k,k'=1,\ldots M-3,
\mid k-k' \mid \geq 2$}  \\
b_k b_{k+1} b_k=b_{k+1} b_k b_{k+1} &
\makebox{      for $k=1,\ldots,M-4$}
\end{array}
\eeq
and
\beq
\tau'(\hat{b})=D^{M-2}\tr(\pi(b))
\eeq
where the trace is normalized as $\tr(1)=1$.
Indeed, omitting to write the configuration $c$ on which the operators act,
by applying repeatedly first and second Markov moves, as well as
the braid group relations, we obtain
\bea
\tau'(\hat{gb_{M-1}})= \nonumber \\
D^{M-1}\tr(\pi(g)T_{2M-1}T_{2M-2}T_{2M}T_{2M-1}^{-1})= \nonumber \\
D^{M-1}\tr(T_{2M-1}^{-1} \pi(g) T_{2M-1}T_{2M-2}T_{2M})= \nonumber \\
D^{M-3/2}\tr(T_{2M-1}^{-1} \pi(g) T_{2M-1}T_{2M-2})= \\
D^{M-3/2}\tr(\pi(g) T_{2M-1}T_{2M-2}T_{2M-1}^{-1})= \nonumber \\
D^{M-3/2}\tr(\pi(g) T_{2M-2}^{-1}T_{2M-1}T_{2M-2})= \nonumber \\
D^{M-3/2}\tr(T_{2M-2} \pi(g) T_{2M-2}^{-1}T_{2M-1})= \nonumber \\
D^{M-2}\tr(T_{2M-2} \pi(g) T_{2M-2}^{-1})= \nonumber \\
\tau'(\hat{g}) \nonumber
\eea
where $g \epsilon B_{M-1}$,$b_{M-1} \epsilon B_M$.
To summarize, we have shown that the ordinary trace on the $Y_{k0}$
is invariant under the Markov moves 1 and 2, and hence provides tangle
invariants (The tangles are in correspondence with the Yang-Baxter operators).
We shall collect the results of this section in table~\ref{tab}.

\begin{table}[t]
\begin{picture}(200,500)(45,-225)
\put(0,0){
\begin{tabular}{|r|lllll|l|l|} \hline
 & & & & & Rapidities & Boltzmann Weights & Yang-Baxter Operators\\
 & $\frac{p}{q}$ & $\frac{q'}{p}$ & $\frac{p'}{q'}$ & $\frac{p'}{q}$ &
 (Spectral Limits) & (IRF-type R-matrix) & (Braid Group Generators)\\
 \hline \hline
 Ia) & 0 & $\infty$ & 1 & $\infty$ & $p \ll q \ll p'=q'$ &
$\omega^{B(\delta,\beta)-B(\delta,\delta)+B(\alpha,\delta-\beta)}$
& $T_k$ $c(k)=1$ $\forall k$ \\
\hline
 Ib) & $\infty$ & $\infty$ & 1 & $\infty$ & $q \ll p \ll p'=q'$ &
$\omega^{-B(\beta,\delta)+B(\beta,\beta)+B(\alpha,\delta-\beta)}$
& $T_k^{-1}$ $c(k)=1$ $\forall k$ \\
\hline
IIa) & 1 & 0 & 0 & 0 & $p' \ll q' \ll p=q$ &
$\omega^{B(\beta,\delta)-B(\beta,\beta)-B(\gamma,\delta-\beta)}$
& $T_k$ $c(k)=-1$ $\forall k$ \\
\hline
IIb) & 1 & 0 & $\infty$ & 0 & $q' \ll p' \ll p=q$ &
$\omega^{-B(\delta,\beta)+B(\delta,\delta)-B(\gamma,\delta-\beta)}$
& $T_k^{-1}$ $c(k)=-1$ $\forall k$ \\
\hline
IIIa) & 1 & $\infty$ & 0 & $\infty$ & $p=q \ll p' \ll q'$ &
$\omega^{B(\beta,\delta)-B(\delta,\delta)+B(\delta-\beta,\gamma)}$
& $T_k$ $c(k)=1$ $\forall k$ \\
\hline
IIIb) & 1 & $\infty$ & $\infty$ & $\infty$ & $p=q \ll q' \ll p'$ &
$\omega^{-B(\delta,\beta)+B(\beta,\beta)+B(\delta-\beta,\gamma)}$
& $T_k^{-1}$ $c(k)=1$ $\forall k$ \\
\hline
IVa) & 0 & 0 & 1 & 0 & $p'=q' \ll p \ll q$ &
$\omega^{B(\delta,\beta)-B(\beta,\beta)-B(\delta-\beta,\alpha)}$
& $T_k$ $c(k)=-1$ $\forall k$ \\
\hline
IVb) & $\infty$ & 0 & 1 & 0 & $p'=q' \ll q \ll p$ &
$\omega^{-B(\beta,\delta)+B(\delta,\delta)-B(\delta-\beta,\alpha)}$
& $T_k^{-1}$ $c(k)=-1$ $\forall k$ \\
\hline \hline
Va) & 0 & 0 & 0 & 0 & $p' \ll q' \ll p \ll q$ &
$\sum_{\mu \epsilon L} \omega^{B(\beta-\gamma,\alpha-\beta)}$
& $T_{2k+1}T_{2k}T_{2k+2}T_{2k+1}^{-1}$ \\
& & & & & & $\omega^{-B(\mu,\mu)+B(\mu,\beta+\delta-\alpha-\gamma)}$ & \\
\hline
Vb) & $\infty$ & 0 & $\infty$ & 0 & $q' \ll p' \ll q \ll p$ &
$\sum_{\mu \epsilon L} \omega^{-B(\delta-\gamma,\alpha-\delta)}$
& $T_{2k+1}T_{2k}^{-1}T_{2k+2}^{-1}T_{2k+1}^{-1}$ \\
& & & & & & $\omega^{B(\mu,\mu)+B(\mu,\beta+\delta-\alpha-\gamma)}$ & \\
\hline
VIa) & 0 & $\infty$ & 0 & $\infty$ & $p \ll q \ll p' \ll q'$ &
$\sum_{\mu \epsilon L} \omega^{B(\delta-\alpha,\gamma-\delta)}$
& $T_{2k+1}^{-1}T_{2k}T_{2k+2}T_{2k+1}$ \\
& & & & & & $\omega^{-B(\mu,\mu)+B(\mu,\beta+\delta-\alpha-\gamma)}$ & \\
\hline
VIb) & $\infty$ & $\infty$ & $\infty$ & $\infty$ & $q \ll p \ll q' \ll p'$ &
$\sum_{\mu \epsilon L} \omega^{-B(\beta-\alpha,\gamma-\beta)}$
& $T_{2k+1}^{-1}T_{2k}^{-1}T_{2k+2}^{-1}T_{2k+1}$ \\
& & & & & & $\omega^{B(\mu,\mu)+B(\mu,\beta+\delta-\alpha-\gamma)}$ & \\
\hline
VIIa) & 0 & $\infty$ & $\infty$ & $\infty$ & $p \ll q \ll q' \ll p'$ &
$D \prod_{i=1}^n\delta_{\beta_{i+1}-\gamma_{i+1},\alpha_i-\delta_i}$
& $T_{2k+1}^{-1}T_{2k}T_{2k+2}^{-1}T_{2k+1}$ \\
& & & & & & $\omega^{B(\delta,\gamma)-B(\alpha,\beta)}$ & \\
\hline
VIIb) & $\infty$ & $\infty$ & 0 & $\infty$ & $q \ll p \ll p' \ll q'$ &
$D \prod_{i=1}^n\delta_{\delta_{i+1}-\gamma_{i+1},\alpha_i-\beta_i}$
& $T_{2k+1}^{-1}T_{2k}^{-1}T_{2k+2}T_{2k+1}$ \\
& & & & & & $\omega^{B(\beta-\alpha,\beta-\delta)}$ & \\
\hline
VIIIa) & 0 & 0 & $\infty$ & 0 & $q' \ll p' \ll p \ll q$ &
$D \prod_{i=1}^n\delta_{\beta_{i+1}-\alpha_{i+1},\gamma_i-\delta_i}$
& $T_{2k+1}T_{2k}T_{2k+2}^{-1}T_{2k+1}^{-1}$ \\
& & & & & & $\omega^{B(\delta-\gamma,\delta-\beta)}$ & \\
\hline
VIIIb) & $\infty$ & 0 & 0 & 0 & $p' \ll q' \ll q \ll p$ &
$D \prod_{i=1}^n\delta_{\delta_{i+1}-\alpha_{i+1},\gamma_i-\beta_i}$
& $T_{2k+1}T_{2k}^{-1}T_{2k+2}T_{2k+1}^{-1}$ \\
& & & & & & $\omega^{B(\beta,\alpha)-B(\gamma,\delta)}$ & \\
\hline
IXa) & 0 & $\infty$ & 0 & 0 & $p \ll p' \ll q \ll q'$ &
$D \delta_{\delta-\gamma,\alpha-\beta}
\omega^{B(\alpha-\delta,\delta-\gamma)}$
& $T_{2k+1}T_{2k}T_{2k+2}T_{2k+1}$ \\
Xa) & & & & & $p' \ll p \ll q' \ll q$ & & \\
\hline
IXb) & $\infty$ & 0 & $\infty$ & $\infty$ & $q \ll q' \ll p \ll p'$ &
$D \delta_{\delta-\gamma,\alpha-\beta}
\omega^{-B(\alpha-\beta,\beta-\gamma)}$
& $T_{2k+1}^{-1}T_{2k}^{-1}T_{2k+2}^{-1}T_{2k+1}^{-1}$ \\
Xb) & & & & & $q' \ll q \ll p' \ll p$ & & \\
\hline
\end{tabular}}
\end{picture}
\caption{The spectral limits and the resulting Boltzmann weights
and Yang-Baxter operators}
\label{tab}
\end{table}

\newsection{Generalizations}

In the previous section we have shown that the 3-D-Baxter-Bazhanov model
can be related to the cyclotomic knot invariants generated by the limits
I)-IV) of the associate Yang-Baxter operators $Y_k$. Under the other limits
V)-X) one obtains products like
\beq
T_{\pm 2k+1}(s_{2k+1}(c))T_{\pm 2k}(s_{2k+1}(c))
T_{\pm 2k+2}(s_{2k+1}(c))T_{\pm 2k+1}(c)
\label{prop}
\eeq
It is intriguing to think that the products~(\ref{prop})
give a "cabling" representation of the braid group, analogously
to the procedure established by Akutsu, Wadati, Deguchi~\cite{aku1,aku2} to
construct higher-dimensional braid group representation
of the Hecke algebra of $B_M$. However some observations are in order:

i) The cabling here is, perhaps, related to higher-dimensional
representations of $U_q(\hat{gl}(n))$ with $q^N=1$.

ii) Probably we must give up the orientation, and hence the invariants are
of non-oriented type.

iii) The single $T_k$ are related to a representation of the
Temperley-Lieb algebra~\cite{kobaiashi,gold}
only for N=2,3, n=2. Therefore only in these cases one may think to
generalize the construction implemented in \cite{aku1,aku2}.

Work along this direction is in progress.

\vspace{20pt}

\begin{center}
{\bf Aknowledgements}
\end{center}
We would like to thank Paolo Cotta Ramusino for some helpful discussions.

\end{document}